\documentclass[onecolumn,nofootinbib,preprintnumbers, sort&compress, floatfix,a4paper,superscriptaddress,10pt]{revtex4-1}
\usepackage[framemethod=TikZ]{mdframed}
\usepackage[most]{tcolorbox}
\usepackage{longtable}
\usepackage{mathcomp}
\usepackage{pifont}
\usepackage{xcolor}
\usepackage{amsthm}
\usepackage{pgf}
\usepackage{pgfpages}
\usepackage{amsmath}
\usepackage{amsfonts}
\usepackage{amssymb}
\usepackage{graphicx}
\usepackage{bm}
\usepackage{hyperref}
\usepackage{lipsum}
\usepackage{array}
\usepackage{booktabs}
\usepackage{tikz}
\usepackage{titlesec}
\usepackage{hyperref}
\hypersetup{bookmarksnumbered}
\def\be{\begin{equation}}
\def\ee{\end{equation}}

\usepackage{amsmath}
\usepackage{empheq}
\usepackage[most]{tcolorbox}
\newtcbox{\mymath}[1][]{%
    nobeforeafter, math upper, tcbox raise base,
    enhanced, colframe=blue!30!black,
    colback=blue!30, boxrule=1pt,
    #1}
\usepackage{mdframed}
\usepackage{lipsum}
\usepackage{PGF}
\usepackage{pgfkeys}
\usepackage{pgffor}
\usepackage{pgfcalendar}
\usepackage{pgfpages}
\usepackage{tikz}
\usetikzlibrary{calc}
\usepackage{array}
\usepackage{booktabs}
\usepackage{multirow}
\usepackage{latexsym}
\usepackage[english]{babel}
\allowdisplaybreaks
\vbox{\hbox{1000}}
\hypersetup{colorlinks=true,linkcolor=blue!100,citecolor=red!100}
\newcounter{theo}[section] 
\renewcommand{\thetheo}{\arabic{section}.\arabic{theo}}

\newcounter{prf}[section]
\renewcommand{\theprf}{\arabic{section}.\arabic{prf}}

\newcounter{lem}[section]
\renewcommand{\thelem}{\arabic{section}.\arabic{lem}}

\titleformat{\section}
{\color{black}\normalfont\Large\bfseries}
{\color{black}\thesection}{1em}{}
\renewcommand\thesection{\arabic{section}}

\newcommand*{\hrmybox}[2]{\colorbox[rgb]{0.70,0.70,1.00}{\parbox{.99\linewidth}{#2}}}
\newcommand*{\rmybox}[2]{\colorbox[rgb]{0.91,0.45,0.45}{\parbox{.99\linewidth}{#2}}}

\newcommand*{\pqrmybox}[2]{\colorbox[rgb]{0.00,0.98,0.25}{\parbox{.99\linewidth}{#2}}}
\newcommand*{\goldmybox}[2]{\colorbox[rgb]{0.76,0.98,0.25}{\parbox{.99\linewidth}{#2}}}

\newcommand*{\bahbahmybox}[2]{\colorbox[rgb]{1.00,0.42,1.00}{\parbox{.99\linewidth}{#2}}}
\allowdisplaybreaks
\usepackage{booktabs}

\begin{document}
\title{\textcolor[rgb]{0.00,0.40,0.80}{\textsf{\huge\boldmath
Raychaudhuri-based reconstruction of anisotropic Einstein-Maxwell equation in 1+3 covariant formalism of \(f(R)\)-gravity}}}
\author{‪Behzad Tajahmad}
\email{behzadtajahmad@yahoo.com}
\affiliation{Faculty of Physics, University of Tabriz, Tabriz, Iran}
\affiliation{Research Institute for Astronomy and Astrophysics of Maragha (RIAAM)-Maragha, Iran, P.O. Box: 55134-441\\}
\begin{center}
\begin{abstract}
\begin{tcolorbox}[breakable,colback=white,
colframe=cyan,width=\dimexpr\textwidth+0mm\relax,enlarge left by=-17mm,enlarge right by=-6mm ]
\text{ }\\
\large{\textbf{\textsf{Abstract:}} Recently, a new strategy to the reconstruction of \(f(R)\)-gravity models based on the Raychaudhuri equation has been suggested by Choudhury et al. \cite{gupta}.
In this paper, utilizing this method, the reconstruction of anisotropic Einstein-Maxwell equation in the $1+3$ covariant formalism of \(f(R)\)-gravity is investigated in four modes: \textit{i.} Reconstruction from a negative constant deceleration parameter refereeing to an ever-accelerating universe; \textit{ii.} Reconstruction from a constant jerk parameter $j=1$ which recovers celebrated $\Lambda \text{CDM}$ mode of evolution; \textit{iii.} Reconstruction from a variable jerk parameter $j=Q(t)$; and \textit{iv.} Reconstruction from a slowly varying jerk parameter. \\
Furthermore, two suggestions for enhancing the method are proposed.}
\end{tcolorbox}
\end{abstract}

\end{center}
\maketitle
\hrule \hrule \hrule
\textbf{\textcolor[rgb]{0.00,0.00,0.00}{\tableofcontents}}
\text{ }
\hrule \hrule \hrule
%
\noindent\hrmybox{}{\section{Introduction\label{sec:1}}}\vspace{5mm}

Various astronomical and cosmological observations suggest that the present universe is undergoing a period of an accelerated expansion directly \cite{02,03,04} and indirectly \cite{05,06,07}.\\

It is argued that an unknown hiddenly characteristics sort of energy with large negative pressure is responsible for this accelerating cosmic expansion \cite{08}. This mysterious candidate being incompatible with strong energy condition is dubbed as dark energy.
This cosmic behavior can also be explained either by modifying matter part ($f(R)$ gravity, $f(T)$ gravity, $f(T)$ gravity with unusual term \cite{behinter}, scalar-tensor theories \cite{review}, etc.) or by modifying geometric part (chaplygin gas \cite{chaplygin}, quintessence \cite{quintessence}, phantom \cite{phantom}, quintom \cite{quintom}, etc.) of the Einstein-Hilbert action. This new set of gravity theories passes several solar system and astrophysical tests successfully \cite{od1,od2}.\\

The simple modification of Einstein's theory of gravity namely $f(R)$-gravity, as a source of acceleration, was proposed by  Capozziello et al.~\cite{capp1} and Carroll et al.~\cite{carloni}.
The $f(R)$ model gives sufficient generality to encapsulate some of the basic characteristics of higher-order gravity and yet are rather simple to handle.
The modified $f(R)$-gravity can elucidate the cosmic acceleration without introducing the dark energy component \cite{capp2,behsanyal,early}. In addition,
it has been demonstrated that the modified $f(R)$-gravity can be derived from string/M-theory \cite{string}.\\

One of the unsolved problems in cosmology is the cosmological magnetic field emerging at large scale of the universe observationally \cite{mag1}. In order to disentangle the origin of the primordial cosmological magnetic field, there are many theoretical explanations, for example, it has been created from the Big Bang like all matters populating the universe \cite{prio}. Besides this, such fields might play some roles on the cosmic microwave background radiation.
For these reasons, in this paper, the primordial magnetic fields are included in the energy-momentum tensor of the Einstein field equation directly.
It is worthwhile mentioning that the cosmological magnetic fields will naturally appear in the universe when the anisotropic cosmological models are taken into account. Hence we would like to consider the problem in an anisotropic background.\\

Cosmological models are investigated via several powerful tools: exact solutions, phase space \cite{phas1,phas2}, $\mathfrak{B}$-function method \cite{bfm}, Noether symmetry approach \cite{nod1,nod2,nod3,nod4}, Beyond Noether symmetry approach (B.N.S.) \cite{b.n.s.}, Noether symmetry approach using CSSS-trick \cite{csss}, Reconstruction methods \cite{rec1,rec5,rec2,behrec,rec3}, etc.
Recently, a new strategy to the reconstruction of $f(R)$-gravity models using Raychaudhuri equation has been proposed by Choudhury et al. \cite{gupta}.
In this paper, from the perspective of this new method, we investigate the reconstruction of anisotropic Einstein-Maxwell equation in 1+3 covariant formalism of $f(R)$-gravity in four modes of evolution: \textit{i.} Reconstruction from a negative constant deceleration parameter refereeing to an ever-accelerating universe; \textit{ii.} Reconstruction from a constant jerk parameter $j=1$ which recovers celebrated $\Lambda \text{CDM}$ mode of evolution; \textit{iii.} Reconstruction from a time-variable jerk parameter $j=Q(t)$; and \textit{iv.} Reconstruction from a slowly varying jerk parameter. The two last jerk types for $f(T)$-gravity has recently been studied by Chakrabarti et al. \cite{hhh3}. Finally, some suggestions for enhancing the method are proposed.\\

\noindent\hrmybox{}{\section{The model and basic equations\label{sec:2}}}\vspace{5mm}

In this section, the evolution equations of the $f(R)$-gravity in orthogonally spatially homogeneous 1+3 covariant approach have been set up.

For a given fluid four-velocity vector field $u^{\mu}$, the projection tensor $h_{\mu \nu}=g_{\mu \nu}+u_{\mu}u_{\nu}$ projects into the  instantaneous rest-space of a comoving observer who, in this paper, is characterized by $u^{\mu}=(1,0,0,0)$. Indeed, the four-velocity $u^{\mu}$ is orthogonal to the induce metric $h_{\mu \nu}$ (i.e. $h_{\mu \nu}u^{\mu}=0$).

Introducing the vorticity tensor $\omega_{\mu \nu}$ ($\omega_{\mu \nu}=\omega_{[\mu \nu]}$, $\omega_{\mu \nu}u^{\nu}=0$), the symmetric shear tensor $\sigma_{\mu \nu}$ ($\sigma_{\mu \nu}=\sigma_{(\mu \nu)}$, $\sigma_{\mu \nu}u^{\nu}=0$, ${\sigma^{\alpha}}_{\alpha}=0$), and the volume expansion $\Theta=\nabla_{\alpha}u^{\alpha}$ the first covariant derivative of the four-velocity can therefore be decomposed as
\begin{equation}
    \nabla_{\mu}u_{\nu}=-u_{\mu}\dot{u}_{\nu}+\omega_{\mu \nu}+\sigma_{\mu \nu}+\frac{1}{3}\Theta h_{\mu \nu},
	\label{firstde}
\end{equation}
where $\dot{u}_{\mu}$, which is defined as $\dot{u}_{\mu}=u^{\nu}\nabla_{\nu}u_{\mu}$, is the acceleration vector. The last term in~(\ref{firstde}) is indeed the following difference
\begin{equation}
    \frac{1}{3}\Theta h_{\mu \nu}=\Theta_{\mu \nu}-\sigma_{\mu \nu},
	\label{the1}
\end{equation}
where $\Theta_{\mu \nu}$ are the components of the volume expansion tensor of the fluid (or the extrinsic curvature) whose its trace (i.e. $\Theta \equiv \Theta_{\mu \nu} h^{\mu \nu}$) is the rate of the volume expansion parameter namely Hubble parameter.

Relatively to $u^{\mu}$, the energy-momentum tensor can be decomposed in the form:
\begin{equation}
    T_{\mu \nu}=\rho u_{\mu} u_{\nu}+2u_{(\nu}q_{\mu)}+p h_{\mu \nu}+\pi_{\mu \nu},
	\label{the2}
\end{equation}
where $\rho$ is the energy density, $p$ is the isotopic pressure, $q_{\mu}$ is the energy flux ($q_{\alpha}u^{\alpha}=0$), and $\pi_{\mu \nu}$ is the symmetric trace-free anisotropic stress pressure ($\pi_{\mu \nu}=\pi_{\nu \mu}$, ${\pi^{\alpha}}_{\alpha}=0$, $\pi_{\mu \nu}u^{\mu}=0$).

We start with the gravitational action of \(f(R)\)-gravity of the form:
\begin{equation}
    S=\int \sqrt{-g}f(R) d^{4}x + \int \mathcal{L}_{\mathcal{M}} d^{4}x,
	\label{action}
\end{equation}
where $g$ is the determinant of metric, $f(R)$ is a function of the Ricci scalar $R$, and $\mathcal{L}_{\mathcal{M}}$ stands for the matter fields Lagrangian density. Varying this action with respect to metric tensor $g^{\mu \nu}$, the Einstein field equations are obtained as follows\footnote{We use natural units: $c=8 \pi G =\hbar =1$. Also, note that $T^{\mathcal{M}}_{\mu \nu} \equiv (2/\sqrt{-g})\delta \mathcal{L}_{\mathcal{M}}/\delta g^{\mu \nu}$.}:
\begin{equation}
    f^{\prime}(R) R_{\mu \nu}-\frac{1}{2}f(R)g_{\mu \nu}+g_{\mu \nu} \nabla_{\alpha} \nabla^{\alpha}f^{\prime}(R)-\nabla_{\mu} \nabla_{\nu}f^{\prime}(R)=T^{\mathcal{M}}_{\mu \nu},
	\label{fe1}
\end{equation}
where $f^{\prime}(R)=\mathrm{d}f/\mathrm{d}R$, and all the subscripts and superscripts run from zero to three (i.e. $0$, $1$, $2$, and $3$). Assuming the total energy-momentum tensor, $T^{\mathcal{M}}_{\mu \nu}$, consists of an electromagnetic filed, $T^{\mathrm{em}}_{\mu \nu}$, and a perfect fluid, $T^{\mathrm{pf}}_{\mu \nu}$, as two main non-interacting parts, it can therefore be written as
\begin{equation}
    T^{\mathcal{M}}_{\mu \nu}=T^{\mathrm{pf}}_{\mu \nu}+T^{\mathrm{em}}_{\mu \nu},
	\label{Tab}
\end{equation}
where\footnote{Note that the subscript $\mathrm{m}$ does not refer to dust matter only, but it stands for perfect fluid matter in general.}
\begin{equation}
    T^{\mathrm{pf}}_{\mu \nu}=\rho_{m} u_{\mu} u_{\nu}+p_{\mathrm{m}}h_{\mu \nu},
	\label{Tab1}
\end{equation}
and
\begin{equation}
    T^{\mathrm{em}}_{\mu \nu}=F_{\mu \alpha} {F^{\alpha}}_{\nu}-\frac{1}{4}g_{\mu \nu} F_{\alpha \beta}F^{\alpha \beta},
	\label{Tab2}
\end{equation}
in which $F_{\mu \nu}$ is the field strength. For a given electric field, $E_{\mu}$, and magnetic field, $B_{\mu}$, the field strength $F_{\mu \nu}$ is defined as
\begin{equation}
    F_{\mu \nu}=\frac{1}{2}u_{[\mu}E_{\nu]}+\eta_{\mu \nu \alpha \beta}B^{\alpha}u^{\beta},
	\label{Tab3}
\end{equation}
in which $\eta_{\mu \nu \alpha \beta}$ is an antisymmetric permutation tensor of space-time with $\eta_{0123}=\sqrt{-g}$. The energy-momentum tensor of Maxwell field can be recast in the form
\begin{equation}
    T^{\mathrm{em}}_{\mu \nu}=\rho_{\mathrm{em}}u_{\mu} u_{\nu}+p_{\mathrm{em}} h_{\mathrm{em}}+\pi_{\mathrm{em}},
	\label{Tab4}
\end{equation}
where $\rho_{\mathrm{em}}$ and $p_{\mathrm{em}}$ are the energy density and the isotropic pressure of the electromagnetic field, respectively, and they are given by
\begin{equation}
    \rho_{\mathrm{em}}=3 p_{\mathrm{em}}= \frac{1}{3} \left(E^2 + B^2 \right),
	\label{Tab5}
\end{equation}
and the anisotropic stress is
\begin{equation}
    \pi_{\mu \nu}= -E_{\mu}E_{\nu}-B_{\mu}B_{\nu}+\frac{1}{3}\left(E^2 + B^2 \right)h_{\mu \nu}.
	\label{Tab6}
\end{equation}
In the present paper, we prefer to work with a pure magnetic case (i.e. $E=0$ and $B \neq 0$).\\
Let us consider the problem in the anisotropic background of the form
\begin{equation}
    \mathrm{d}s^2 = -\mathrm{d}t^2 + a^2(t) \; \mathrm{d}x^2 + b^2(t) \left( \mathrm{d}y^2 + \mathrm{d}z^2 \right).
	\label{metric}
\end{equation}
This line element is known as Locally Rotationally Symmetric Bianchi type-I (LRS B-I). Pursuant to this background geometry, the magnetic fields may have component as $B_{\mu}=(0; B(t), 0, 0)$.

Defining $S_{\mu \nu}=\nabla_{\mu} \nabla_{\nu}f^{\prime}$ and using equation~(\ref{fe1}), the Ricci tensor takes the following form
\begin{equation}
    R_{\mu \nu}=\frac{1}{f^{\prime}}\left(\frac{1}{2}g_{\mu \nu}f - \left(g_{\mu \nu}g^{\alpha \beta} - g^{\alpha}_{\mu}g^{\beta}_{\nu} \right)S_{\alpha \beta}+ T^{\mathcal{M}}_{\mu \nu} \right).
	\label{ricci}
\end{equation}
Now, utilizing this equation, the Ricci tensor $R_{\mu \nu}$ can be split into the following forms:
\begin{align}
    &R=\frac{1}{f^{\prime}} \left(T^{\mathcal{M}}+2f-3S \right),
	\label{ricci1}\\
    &R_{\mu \nu}u^{\mu} u^{\nu}=\frac{1}{f^{\prime}} \left(T^{\mathcal{M}}_{\mu \nu}u^{\mu} u^{\nu} -\frac{1}{2}f+h^{\mu \nu}S_{\mu \nu} \right),
	\label{ricci2}\\
    &R_{\mu \nu}u^{\mu} h^{\nu}_{\alpha}=\frac{1}{f^{\prime}} \left(S_{\mu \nu}u^{\mu}h^{\nu}_{\alpha} - q_{\alpha} \right),
	\label{ricci3}\\
    &R_{\mu \nu}h^{\mu}_{\alpha} h^{\nu}_{\beta}=\frac{1}{f^{\prime}} \left(S_{\mu \nu}h^{\mu}_{\alpha} h^{\nu}_{\beta}+\pi_{\alpha \beta} - \left(p+S+\frac{1}{2}f \right) \right),
	\label{ricci4}
\end{align}
where $p=p_{\mathrm{tot.}}=p_{\mathrm{m}}+p_{\mathrm{em}}$. In analogous with the Ricci tensor, for the $S_{\mu \nu}$ one has the following relations:
\begin{align}
    &S=-f^{\prime \prime} \left(\ddot{R}+\Theta \dot{R} \right)- f^{\prime \prime \prime} \dot{R}^{2},
	\label{S1}\\
    &S_{\mu \nu} u^{\mu} u^{\nu}=f^{\prime \prime} \ddot{R}+f^{\prime \prime \prime} \dot{R}^{2},
	\label{S2}\\
    &S_{\mu \nu} h^{\mu \nu}=-f^{\prime \prime} \Theta \dot{R}.
	\label{S3}
\end{align}
Therefore, it can be demonstrated that the Raychaudhuri equation in the 1+3 covariant formalism of $f(R)$-gravity of Bianchi type-I is obtained as
\begin{equation}
    \dot{\Theta}+\frac{1}{3}\Theta^{2}+2 \sigma^{2}+\frac{1}{f^{\prime}}\left(\rho-\frac{1}{2}f-f^{\prime \prime}\Theta \dot{R} \right)=0,
	\label{Rech1}
\end{equation}
where $\rho=\rho_{\mathrm{tot.}}=\rho_{\mathrm{m}}+\rho_{\mathrm{em}}$. Restricting the magnetic field to be aligned along the shear eigenvector, the shear tensor would also diagonal $\sigma_{\mu \nu}=\mathrm{diag}(\sigma_{11},\sigma_{22},\sigma_{33})$. Pursuant to our background geometry of study~(\ref{metric}), it is comfortably deduced that $\sigma_{11}=-(\sigma_{22}+\sigma_{33})$\footnote{These types of shears may usually be defined as
\begin{equation*}
    \sigma_{ij}=\mathrm{diag}\left[-\frac{2\sigma_{+}}{H},\frac{\sigma_{+}
    +\sqrt{3}\sigma_{-}}{H},\frac{\sigma_{+}
    -\sqrt{3}\sigma_{-}}{H} \right],
\end{equation*}
where $H$ is the Hubble parameter.}. The propagations of matter parts ($\mathrm{pf}$/$\mathrm{m}$ and $\mathrm{em}$) and each element of the shear tensor, are then given by \footnote{Although we present the equations in general form, in examples excluding the last one, the special form of perfect fluid namely pressureless dust matter (i.e. $w=0$) would be the case of study.}
\begin{align}
&{\dot{\rho}}_{\mathrm{m}}+ \left(1+w\right)\Theta \rho_{\mathrm{m}}=0; \qquad w=p_{\mathrm{m}}/\rho_{\mathrm{m}} \;, \label{pfe1}\\
&\dot{B}+\frac{2}{3}\Theta B+\sigma_{11}B=0, \label{eme1}\\
&{\dot{\sigma}}_{\mu \mu}+\Theta \sigma_{\mu \mu}+ \frac{f^{\prime \prime}}{f^{\prime}}\dot{R}\sigma_{\mu \mu}-\frac{1}{f^{\prime}}\pi_{\mu \mu}=0. \label{shear1}
\end{align}
Using equation~(\ref{Rech1}) and its first integral, it can easily be indicated that
\begin{equation}
    R=2 \dot{\Theta} +\frac{4}{3} \Theta^{2} + 2 \sigma^{2}.
	\label{RTS}
\end{equation}
This well-known relation is used frequently in this paper.\\

\noindent\hrmybox{}{\section{Reconstruction of $f(R)$-gravity\label{sec:b1b2nmk}}}\vspace{5mm}

In flat FRW background (i.e. $\mathrm{d}s^2 = -\mathrm{d}t^2 + a^2(t)_{\mathrm{FRW}}(\mathrm{d}x^2 + \mathrm{d}y^2 + \mathrm{d}z^2)$), writing the Taylor expansion of the scale factor down as
\begin{equation}\begin{split}
a(t)_{\mathrm{FRW}}=a_{\mathrm{FRW0}}\bigg\{&1+H_{0}(t-t_{0})
-\frac{1}{2} q_{0} H^2_{0}(t-t_{0})^2
\\&+\frac{1}{6} j_{0} H^3_{0}(t-t_{0})^3
+\frac{1}{24} s_{0} H^4_{0}(t-t_{0})^4+\cdots \bigg\},
	\label{Taylora}
\end{split}\end{equation}
some dimensionless parameters, namely the deceleration, $q$, jerk, $j$, snap, $s$, and others, are appeared and they are defined as \cite{jjj1,jjj2}
\begin{align}
&q=\frac{-\ddot{a}_{\mathrm{FRW}}}{a_{\mathrm{FRW}}}
\left(\frac{\dot{a}_{\mathrm{FRW}}}{a_{\mathrm{FRW}}} \right)^{-2}, \quad
j=\frac{\dddot{a}_{\mathrm{FRW}}}{a_{\mathrm{FRW}}}
\left(\frac{\dot{a}_{\mathrm{FRW}}}{a_{\mathrm{FRW}}} \right)^{-3},
\nonumber \\
&s=\frac{\ddddot{a}_{\mathrm{FRW}}}{a_{\mathrm{FRW}}}
\left(\frac{\dot{a}_{\mathrm{FRW}}}{a_{\mathrm{FRW}}} \right)^{-4}, \quad \cdots.
	\label{Taylora1}
\end{align}
In~(\ref{Taylora}), $a_{\mathrm{FRW0}}$ and $H_{0}$ are the present values of the scale factor and Hubble parameter, respectively.
These parameters are a focus of interest because their amounts give important knowledge about the universe. In our anisotropic background geometry, LRS B-I, the relation~(\ref{Taylora}) can be regarded as the Taylor expansion of the average scale factor $\bar{a}=(a b^{2})^{1/3}$. Consequently, we may rewrite the dimensionless parameters~(\ref{Taylora1}) using the extrinsic curvature $\Theta$. This make our work easy. Furthermore, in order to reconstruct $f(R)$-gravity models by~(\ref{Rech1}), we need $\Theta$, not $a(t)$ or $b(t)$. Hence, we deal with the rate of volume expansion parameter $\Theta$, not the scale factors.\\
In a spatially homogeneous model the ratio of shear scalar $\sigma$ to the to expansion scalar $\Theta$ is constant: $(\sigma / \Theta)=\mathrm{const.}$. This may compel the following conditions among the directional Hubble parameters $H_{a}$ and $H_{b}$ ($H_{a}$ along $x$ direction while $H_{b}$ along $y$ and $z$ directions) and the expansion scalar $\Theta$:
\begin{equation}
    H_{a}=\epsilon_{1} \Theta, \quad H_{b}=\epsilon_{2} \Theta,
	\label{condition1}
\end{equation}
where $\epsilon_{1}$ and $\epsilon_{2}$ are constant, yielding
\begin{equation}
    \sigma^{2}=\frac{(\epsilon_{1}-\epsilon_{2})^2}{3} \Theta^{2},
	\label{condition2}
\end{equation}
as we expect (because $(\sigma/\Theta)=(|\epsilon_{1}-\epsilon_{2}|/\sqrt{3})=\mathrm{const.}$). This physical relation makes our work easy in the calculations. It is important to mention that the special case $\sigma=0$ implies $\epsilon_{1}=\epsilon_{2}=(1/3)$ meaning an isotropic background --- FRW.\\
Note that according to~(\ref{condition2}), the rate of $\sigma$ and $\Theta$ would be equal since $(\dot{\sigma}/\sigma)=(\dot{\Theta}/\Theta)$.\\
Pursuant to observational data, it has been demonstrated in \cite{dep01,behrec} that $|\epsilon_{1}-\epsilon_{2}| /\sqrt{3}$ is of order $10^{-10}$.\\

\noindent\rmybox{}{\subsection{A constant deceleration parameter (An accelerating universe)}}\vspace{5mm}

As the first example, let us focus on the acceleration epoch of the universe. This feature of the universe can be determined by the deceleration parameter $q$ --- for an accelerating universe $q<0$ and for a decelerating universe $q>0$. For the aforementioned purpose, a constant deceleration parameter as
\begin{equation}
    q=-|m|; \qquad m \in \{(-1,0) \cup (0,+1)\},
	\label{dece1}
\end{equation}
is our start point in this part. Writing the deceleration parameter in terms of the extrinsic curvature as
\begin{equation}
    q=-1-3 \frac{\dot{\Theta}}{\Theta^{2}},
	\label{11dece1}
\end{equation}
and combining it with (\ref{dece1}), yields
\begin{equation}
    \Theta=\left(\frac{3}{1-|m|}\right) \frac{1}{t-t_{0}},
	\label{ch3-2}
\end{equation}
where $t_{0}$ is an integration constant which we must take it as an initial time (i.e. at all times of interest $t_{0}<t$) in order to keep expanding phase of the universe (i.e. if $t_{0}<t$ then we have $\Theta >0$). Using the average scale factor $\bar{a}$, the average Hubble parameter $\bar{H}$, and the $\mathfrak{B}\text{-function}$ \cite{bfm}, one has
\begin{equation}
    \mathfrak{B}[\bar{a},0;\bar{H}(\bar{a})]
    =b^{\frac{\epsilon_{1}+2\epsilon_{2}}{3\epsilon_{2}}} \frac{\dot{\Theta}}{\Theta},
	\label{Bf1}
\end{equation}
for this case. According to~(\ref{dece1}) we get
\begin{equation}
    -1<\mathfrak{B}[\bar{a},0;\bar{H}(\bar{a})]<0.
	\label{Bf2}
\end{equation}
According to ref.~\cite{bfm}, this era is an accelerated era with non-phantom-like regime property. The limited amounts $-1$ and $0$ correspond to the inflection point namely shifting from a decelerated to an accelerated expansion (i.e. an expansion with constant rate) and de Sitter era/expansion, respectively. It is worthwhile to mention that the left bound, $-1$, is equivalent to $m=0$ and $w_{\mathrm{eff.}}=-1/3$ and the right bound, $0$, is equivalent to $|m|=1$ and $w_{\mathrm{eff.}}=-1$. Note that we have used the well-known definition
\begin{equation}
    w_{\mathrm{eff.}}=\frac{p_{\mathrm{eff.}}}{\rho_{\mathrm{eff.}}}.
	\label{Bf3}
\end{equation}
in which $\rho_{\mathrm{eff.}}$ and $p_{\mathrm{eff.}}$ are given by
\begin{equation}
    \rho_{\mathrm{eff.}} = \left[\frac{2\epsilon_{1}\epsilon_{2}
    +\epsilon^{2}_{2}}{\left(\epsilon_{1}+2\epsilon_{2} \right)^{2}} \right] \Theta^{2},
	\label{Bf4}
\end{equation}
\begin{equation}
    -p_{\mathrm{eff.}}= \left[\frac{\epsilon^{2}_{1}+\epsilon_{1}\epsilon_{2}
    +4\epsilon^{2}_{2}}{2\left(\epsilon_{1}+2\epsilon_{2} \right)^{2}} \right] \Theta^{2}+
    \left[\frac{\epsilon_{1}+3 \epsilon_{2}}{2\left(\epsilon_{1}+2\epsilon_{2} \right)} \right] \dot{\Theta},
	\label{Bf5}
\end{equation}
whence we get
\begin{equation}
    w_{\mathrm{eff.}} = \frac{2|m|+1}{-3}.
	\label{Bf6}
\end{equation}
It is worth mentioning that this relation is also deducible from the relation
\begin{equation}
    w_{\mathrm{eff.}} = -1-2\frac{\dot{\Theta}}{\Theta^{2}},
	\label{Bf7}
\end{equation}
and equation~(\ref{ch3-2}).
\begin{figure*}
	\includegraphics[width=7 in]{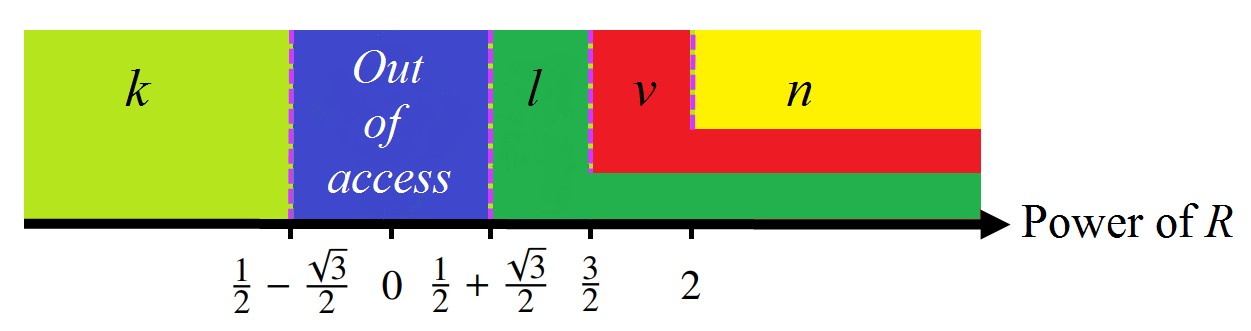}
    \caption{This figure demonstrates the possible orders of $R$ produced by constant parameters.}
    \label{fig1}
\end{figure*}

\begin{figure}
	\includegraphics[width=3.5 in]{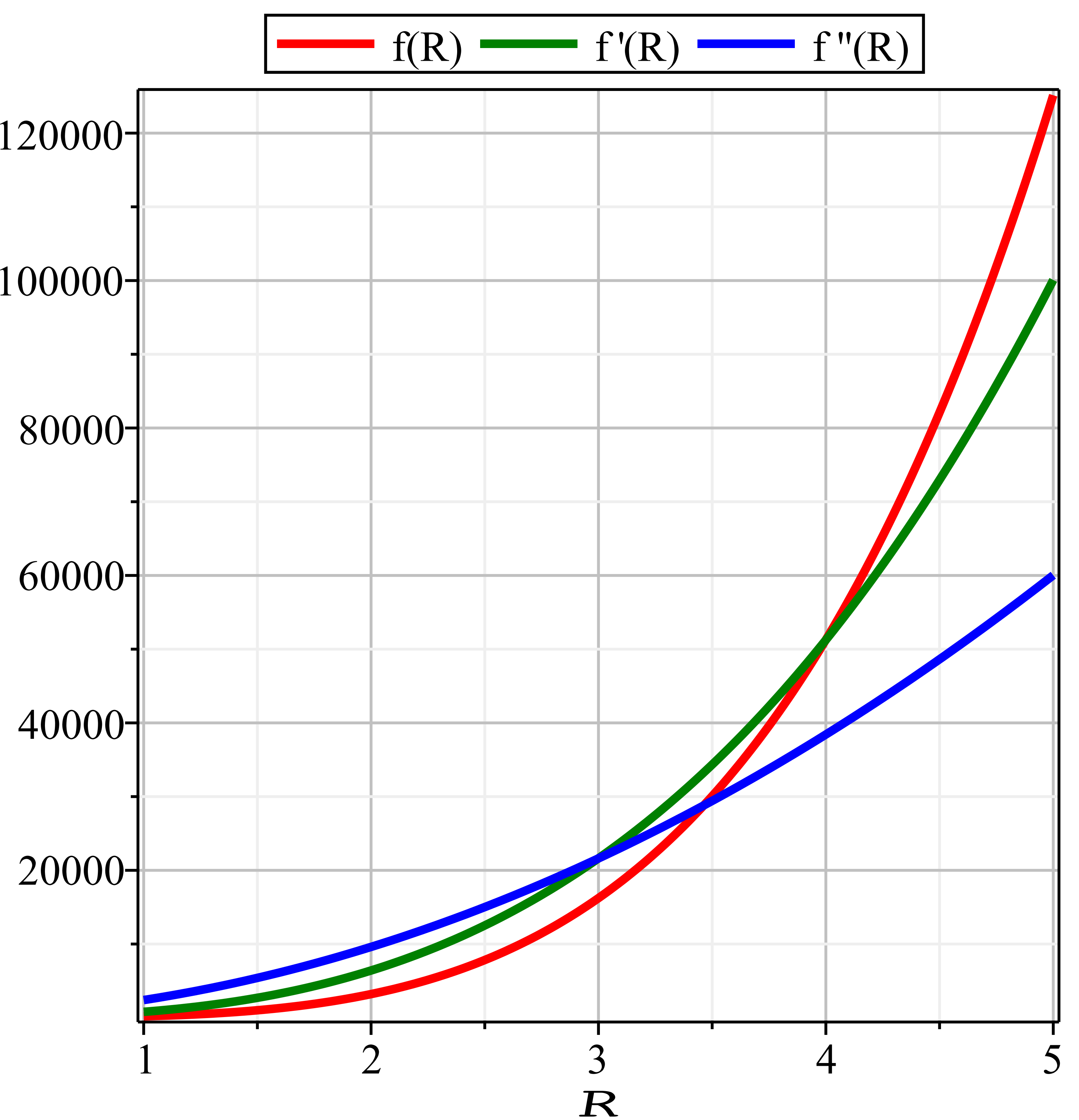}
    \caption{This figure demonstrates the plots of $f(R)$ (red color), $f^{\prime}(R)$ (green color), and $f^{\prime \prime}(R)$ (blue color) for constant deceleration parameter case at curvature interval $[1,5]$.}
    \label{fig1(qc)}
\end{figure}

Using~(\ref{ch3-2}),~(\ref{condition2}), and~(\ref{RTS}) in~(\ref{Rech1}) one arrives at:
\begin{equation}
    R^{2}f^{\prime \prime}+l_{1}Rf^{\prime}+l_{2}f=l_{3}\rho,
	\label{Bf8}
\end{equation}
where
\begin{align}
    l_{1} &=\frac{m}{2(1-m)}+\frac{(\epsilon_{1}-\epsilon_{2})^{2}}{(1-m)},
	\label{Bf9}\\
	l_{2} &=\frac{-3 \gamma}{4(1-m)}, \qquad l_{3}=\frac{-3 \gamma}{2(1-m)},
	\label{Bf10}
\end{align}
in which
\begin{equation}\label{Bf11}
	\gamma =\frac{1}{3}\left\{2(m-1)+2(\epsilon_{1}-\epsilon_{2})^{2}+4 \right\}.
\end{equation}
The total density in (\ref{Bf8}), $\rho=\rho_{\mathrm{pf}}+\rho_{\mathrm{em}}$, is given by
\begin{equation}\label{Bf12}
	\rho=\rho_{\mathrm{m}}+\rho_{\mathrm{em}}
=\underbrace{l_{4}\; R^{v}}_{\mathrm{m-part}}+\underbrace{l_{5}\;R^{n}}_{\mathrm{em-part}}
\end{equation}
where
\begin{align}
	l_{4} &=C_{1} \left(\frac{3 \sqrt{\gamma}}{(1-m)} \right)^{\frac{-3}{1-m}},\label{Bf13}\\
	l_{5} &=\frac{1}{2} C_{2}^{2} \left(\frac{3 \sqrt{\gamma}}{(1-m)} \right)^{\frac{6 \beta}{1-m}},\label{Bf14}\\
	\beta &=\frac{2}{3} \left(1- \frac{\left| \epsilon_{1}-\epsilon_{2} \right|}{\sqrt{3}} \right).\label{Bf15}\\
	v &=\frac{3}{2(1-m)}, \label{Bf16}\\
	n &=\frac{3\beta}{(1-m)}. \label{Bf17}
\end{align}
in which $C_{1}$ and $C_{2}$ are integration constants. Equations~(\ref{Bf8}) and~(\ref{Bf12}) give the form of $f(R)$ as follows:
\begin{equation}\label{Bf18}
f(R)= C_{3} R^{l}+C_{4} R^{k}+l_{8}R^{v}+l_{9}R^{n},
\end{equation}
where $C_{3}$ and $C_{4}$ are integration constants and
\begin{align}
l &=\frac{1-l_{1}}{2}+\frac{\sqrt{l_{1}^{2}-2l_{1}
+1-4l_{2}}}{2},\label{Bf19}\\
k &=\frac{1-l_{1}}{2}-\frac{\sqrt{l_{1}^{2}-2l_{1}
+1-4l_{2}}}{2},\label{Bf20}\\
l_{8} &=\frac{-l_{4}}{v^{2}+(l_{1}-1)v+l_{2}},\label{Bf21}\\
l_{9} &=\frac{-l_{5}}{n^{2}+(l_{1}-1)n+l_{2}}.\label{Bf22}
\end{align}
The terms $l_{8}R^{v}$ and $l_{9}R^{n}$ are related to $\mathrm{m-part}$ and $\mathrm{em-part}$, respectively. In Fig.~\ref{fig1}, the possible powers of $R$ with their generators have been demonstrated for FRW-case. According to~(\ref{Bf15})--(\ref{Bf17}), one always has $n>v$. As is observed, the orders between $(1-\sqrt{3})/2 \approx -0.366$ and $(1+\sqrt{3})/2 \approx + 1.366$ are out of access. The inverse powers of $R$ can be generated only by $k$-parameter. The positive orders of $R$ (here, greater than 1.366) --- so-called higher orders --- can be produced by $l$-parameter. The matter part can only produce the powers greater than $3/2$ while the electromagnetic part can generate greater than $+2$. Therefore, practically, the contribution of both $\mathrm{m-part}$ and $\mathrm{em-part}$ lie on higher orders which can also be generated by $l$-parameter. It means that the feasible orders of $R$ for a vacuum case and a universe filled with matter/electromagnetic/both matter and electromagnetic, are the same. Note that these discussions are valid only for an ever-accelerating universe. It is important to mention that $l$, $k$, and $n$ are affected by anisotropic term, while power $v$, which comes form matter part, is not affected by it. For anisotropic one, the dashed lines in Fig.~\ref{fig1} shift a little (of order $10^{-10}$). Beside all these discussions, it is worthwhile mentioning that there is no way to reach Einstein's theory (i.e. $f(R)=R$), and it may back to this point that Einstein's theory does not lead to an ever-accelerating universe.\\
As is clear, from the mathematical viewpoint, all the integration constants and consequently the parameters $C_{3}$, $C_{4}$, $l_{8}$, and $l_{9}$ can take any complex value in general. Therefore, let us set them to one\footnote{Note that for this purpose, $C_{2}$ must be taken as a pure imaginary number for having $C^{2}_{2}<0$; see equations~(\ref{Bf14}) and~(\ref{Bf22}).} ($C_{3}=C_{4}=l_{8}=l_{9}=1$) for simplicity. It means that the participation amplitude of each term of the obtained form of $f(R)$ in~(\ref{Bf18}) has been normalized to $+1$. In order to compare our plots with the results of ref.~\cite{gupta}, let us take $|m|=0.5$. As is observed from Fig.~\ref{fig1(qc)}, the manner of our plots are different than the ones studied in the aforementioned reference either when we consider limited case namely without $\mathrm{em-part}$ (see figures 2 and 3 in ref.~\cite{gupta}). The plots are presented for FRW case and note that for the anisotropic one, the plots are shifted so little such that they are not visible (i.e. the total behavior will be unchanged). As we know, from the viability analysis viewpoint, two conditions must be adopted for a $f(R)$-model: $f^{\prime}(R)>0$ (for having a positive effective constant of gravitation) and $f^{\prime \prime}(R)>0$ (for the stability of the model). At low curvature, there will be negative and anomalous parts for $f^{\prime}$ and $f^{\prime \prime}$ due to the term $R^{k}$ as $k$ is always negative.
Without loss of generality, one may set $C_{4}=0$ and remove this term, then both validity conditions are satisfied. If one keeps $C_{4}$ non-zero, then there will be a lower bound for the evolution range of curvature. For example, in plotting Fig.~\ref{fig1(qc)}, we kept $C_{4}$, hence there was a lower bound for curvature for the satisfaction of $f^{\prime \prime}>0$ as $R > 0.2411751621$. For this reason, the related plots have been presented for $R \geq 1$. Keeping $C_{4} \neq 0$ causes that the growing speed of $f(R)$ be faster than $f^{\prime}(R)$ and $f^{\prime \prime}(R)$ as $R$ increases; for example, at $R=500$ the values of $f$, $f^{\prime}$, and $f^{\prime \prime}$ are of orders $10^{13}$, $10^{11}$, and $10^{8}$, respectively.\\

\noindent\rmybox{}{\subsection{A constant, a variable, and slowly variable jerk parameters\label{ew15}}}\vspace{5mm}

As is observed from~(\ref{Taylora1}), another interesting dimensionless parameter for focusing is the jerk parameter. It is not hard to show that this parameter in terms of the extrinsic curvature can be written as
\begin{equation}\label{jerk001}
j=9 \frac{\ddot{\Theta}}{\; \Theta^{3}}+9\frac{\dot{\Theta}}{\; \Theta^{2}}+1.
\end{equation}
Three types of jerk values are of physical interest:
\begin{enumerate}
	\item A constant jerk parameter, $j=1$, which mimics the $\Lambda \text{CDM}$ model;
	\item A variable jerk parameter, $j=Q(t)$, such that the jerk parameter is proportional to the Hubble parameter by an inverse square relation. Hence, $Q(t)$ may be taken as $j=Q(t)=\lambda^{2}/H^{2}$ where $\lambda$ is an arbitrary non-zero real constant and $H$ is the mean Hubble parameter.
	\item A slowly varying jerk parameter.
\end{enumerate}
Since the formulation of two first cases of interest is the same in some parts, hence we first consider these two. The slowly varying case will be considered at the end of this section separately.\\
The solutions to equation~(\ref{jerk001}) for $j=1$ and $j=Q(t)$ cases of interest are as follows:
\begin{align}
\text{Sol-I: }\; \Theta_{1} &=n\lambda \tanh [\lambda(t-t_{0}) ],\label{Sol-I}\\
\text{Sol-II: }\; \Theta_{2} &=n\lambda \coth [\lambda(t-t_{0}) ]. \label{Sol-II}
\end{align}
The special values $n=2$ and $n=3$ give $j=1$ and $j=\lambda^{2}/H^{2}$, respectively.
Note that both~(\ref{Sol-I}) and~(\ref{Sol-II}) are a solution to each case with the aforementioned conditions. Therefore, for the first case of interest namely $j=1$, the parameter $\lambda$, in the above solutions, is a free constant parameter while for the second case namely $j=Q(t)$ it is exactly the constant parameter of relation $j=\lambda^{2}/H^{2}$.\\
For~(\ref{Sol-I}) and (\ref{Sol-II}) the effective EoS reads
\begin{align}
w_{\mathrm{eff.1}} &=\frac{p_{\mathrm{eff.1}}}{\rho_{\mathrm{eff.1}}}
=\frac{-2}{n}\coth^{2}[\lambda(t-t_{0})]+\frac{2-n}{n},\label{weff1solI}\\
w_{\mathrm{eff.2}} &=\frac{p_{\mathrm{eff.2}}}{\rho_{\mathrm{eff.2}}}
=\frac{-2}{n}\tanh^{2}[\lambda(t-t_{0})]+\frac{2-n}{n},\label{weff1solII}
\end{align}
respectively. Therefore, for the constant jerk case ($j=1$ or equivalently $n=2$), the boundary values will be
\begin{align}
	&w_{\mathrm{eff.1}}(t \to t_{0})=-\infty, \qquad w_{\mathrm{eff.1}}(t \to \infty)=-1; \label{weff1val1}\\
	&w_{\mathrm{eff.2}}(t \to t_{0})=0, \;\;\;\; \qquad w_{\mathrm{eff.2}}(t \to \infty)=-1,\label{weff2val1}
\end{align}
and for the variable jerk parameter ($j=\lambda^{2}/H^{2}$ or equivalently $n=3$), we have
\begin{align}
	&w_{\mathrm{eff.1}}(t \to t_{0})=-\infty, \;\; \qquad w_{\mathrm{eff.1}}(t \to \infty)=-1; \label{weff1val2}\\
	&w_{\mathrm{eff.2}}(t \to t_{0})=-1/3, \qquad w_{\mathrm{eff.2}}(t \to \infty)=-1.\label{weff2val2}
\end{align}
As is clear from~(\ref{weff1val1}) and~(\ref{weff1val2}), the solution~(\ref{Sol-I}) leads to an unacceptable model because it provides an ever-accelerated universe with wrong behavior of $w_{\mathrm{eff.}}$s because both $w_{\mathrm{eff.}}$s stay in negative region and decay from a high value to $-1$ and consequently it lacks radiation and matter-dominated eras in the past for both jerk types. However, the amounts of $w_{\mathrm{eff.}}$s in (\ref{weff1val1}) and (\ref{weff1val2}), indicate phantom-like regime, but a physical EoS must decrease from positive values (crossing from +1/3 (radiation-dominated era) and 0 (matter-dominated era)) to negative values. Besides these problems, in what follows, it is argued that this type of solution yields an imaginary form for $f(R)$-gravity which is non-physical.
According to~(\ref{weff2val1}) and~(\ref{weff2val2}) the solution~(\ref{Sol-II}) may be called physical as both behaviors of $w_{\mathrm{eff.}}$s are in accordance with a part of the (accelerating) evolution of the observed universe with a non-phantom-like regime. A difference between the two is that $w_{\mathrm{eff.2}}=0$ in~(\ref{weff2val1}) indicates matter-dominated era that pursuant to $w_{\mathrm{eff.2}}=-1$ in~(\ref{weff2val1}), it is followed by an accelerating expansion, while for the next case of study the start point of EoS is $-1/3$ which only refers to an acceleration mode of expansion. This difference may be interpreted as a privilege to $\Lambda \text{CDM}$ model in comparison with a decaying jerk model. Note that whether the obtained forms of $\Theta$ lead to a physical behavior of EoS or not, it is better we study all these cases because they may yield a form of $f(R)$ which may help for giving a model and solving some problems in other subjects of physics (e.g. Inflation). In other words, if an obtained form of $f(R)$ does not work here because of its non-physical outcomes, it may be examined in other models which besides $f(R)$ term there is, for example, a scalar field lagrangian and then it may solve some problems.

For both sets of solutions, (\ref{Sol-I})--(\ref{Sol-II}), the Raychaudhuri equation turns out to be
\begin{align}\label{Ray02}
&\frac{-2}{n}A_{4}\left(R-2n\lambda^{2}\right)^{2}f^{\prime \prime}-2n\lambda^{2}A_{3}\left(R-2n\lambda^{2}\right)f^{\prime \prime} \nonumber \\&+
n \lambda^{2} f^{\prime}+A_{3}\left(R-2n\lambda^{2}\right)f^{\prime}-\frac{1}{2}f+\rho=0
\end{align}
where
\begin{align*}
&A_{3}=A_{1}/A_{2}, \quad A_{4}=A_{3}/A_{2}, \nonumber \\
&A_{1}=\frac{-1}{n}+\frac{1}{3}+\frac{2}{3}(\epsilon_{1}-\epsilon_{2})^2, \nonumber\\
&A_{2}=\frac{-2}{n}+\frac{4}{3}+\frac{2}{3}(\epsilon_{1}-\epsilon_{2})^2.
\end{align*}
The density in~(\ref{Ray02}) is given by
\begin{align}
\rho=\rho_{\mathrm{m}}+\rho_{\mathrm{em}}
\end{align}
in which
\begin{align}
\rho_{\mathrm{m}} &=\rho_{\mathrm{m0}} \left[\epsilon -\epsilon A_{5} \left( R-2n \lambda^{2} \right) \right]^{n/2}, \label{rhom002}\\
\rho_{\mathrm{em}} &=\rho_{\mathrm{em0}} \left[\epsilon -\epsilon A_{5} \left( R-2n \lambda^{2} \right) \right]^{n \left|\beta_{1}\right|},\label{rhoem002}
\end{align}
where $\rho_{\mathrm{m0}}$ and $\rho_{\mathrm{em0}}$ are constants of integration and
\begin{align}
&A_{5}=\left(n^{2} \lambda^{2} A_{2} \right)^{-1}, \\
& \beta_{1}=\frac{2}{3} \left(\frac{\left|\epsilon_{1} -\epsilon_{2} \right|}{\sqrt{3}}-1 \right),\\
&\left\{
  \begin{array}{ll}
    \text{For solution~(\ref{Sol-I})} \; \Leftrightarrow \; \epsilon=+1; \\
    \text{For solution~(\ref{Sol-II})} \; \Leftrightarrow \; \epsilon=-1.
  \end{array}
\right.
\end{align}

\noindent\bahbahmybox{}{\subsubsection{The constant jerk case $j=1$ ($n=2$)}}\vspace{5mm}
\begin{figure}
	\includegraphics[width=3.5 in]{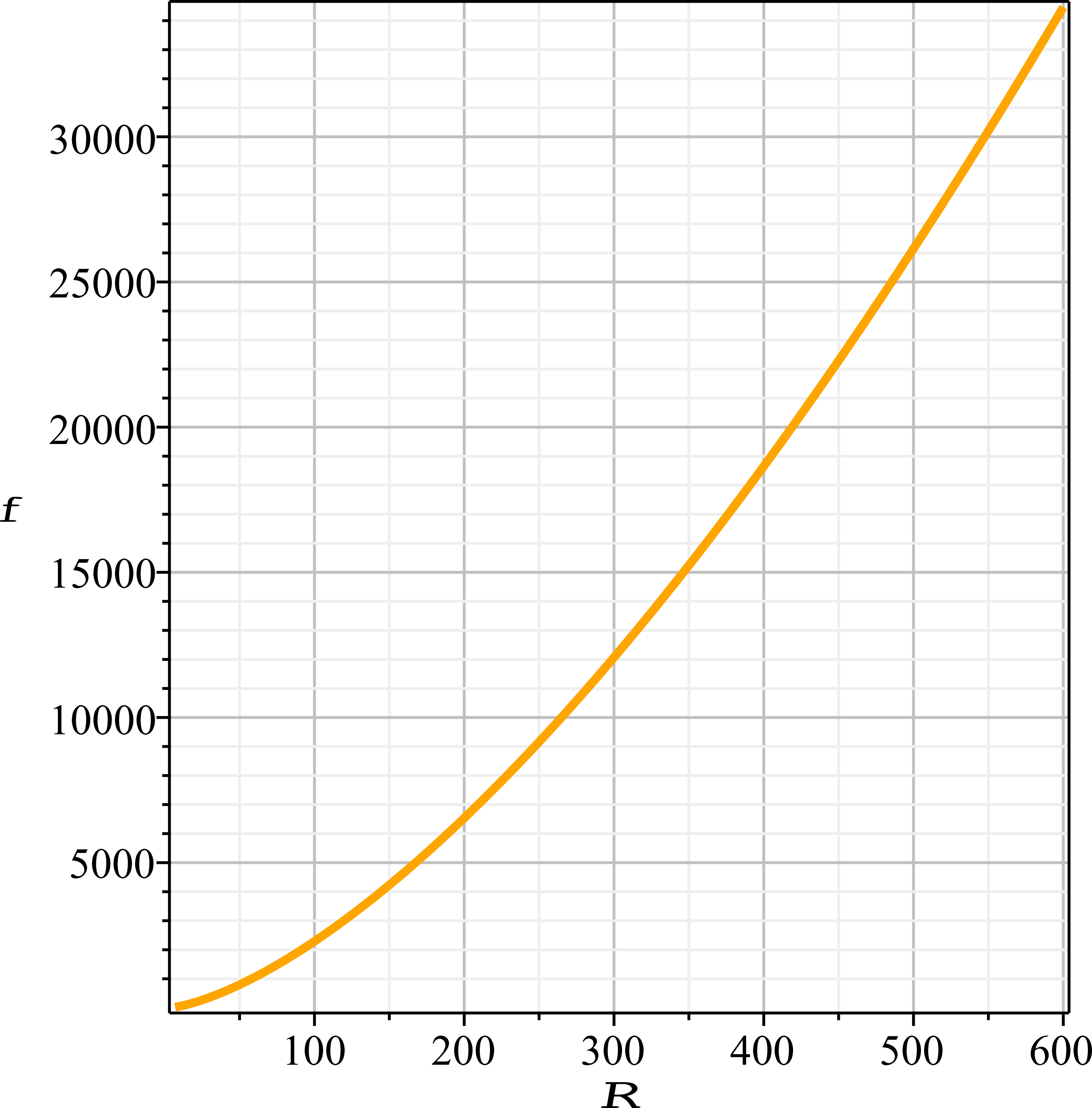}
    \caption{This figure indicates the behavior of $f(R)$ (obtained from numerical analysis) versus curvature $R$ at curvature interval $[7,600]$.}
    \label{f(Num)}
\end{figure}
\begin{figure}
	\includegraphics[width=3.5 in]{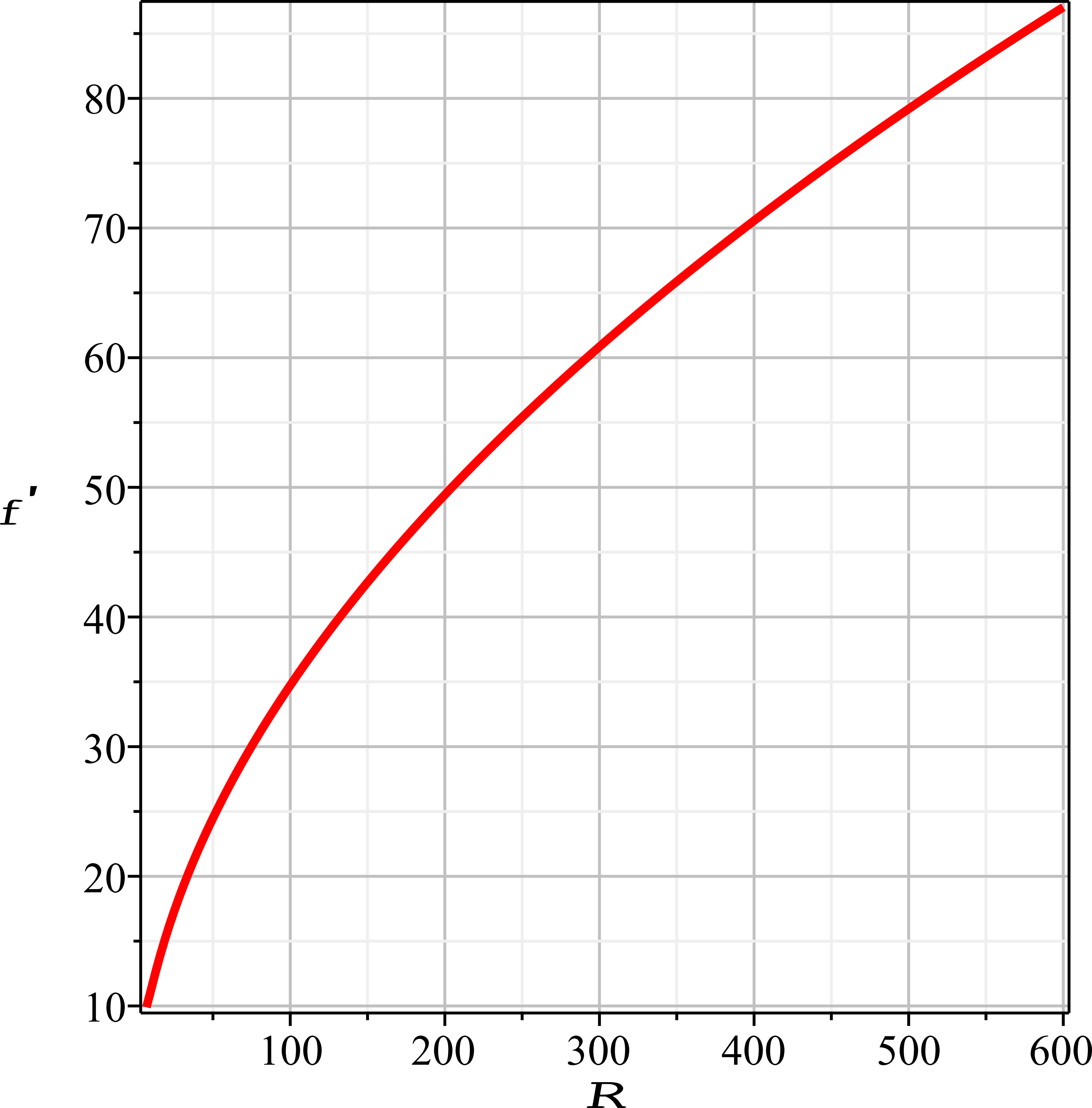}
    \caption{This figure demonstrates the plot of $f^{\prime}(R)$ (obtained from numerical analysis) versus curvature $R$ at curvature interval $[7,600]$.}
    \label{fprime(Num)}
\end{figure}
\begin{figure}
	\includegraphics[width=3.5 in]{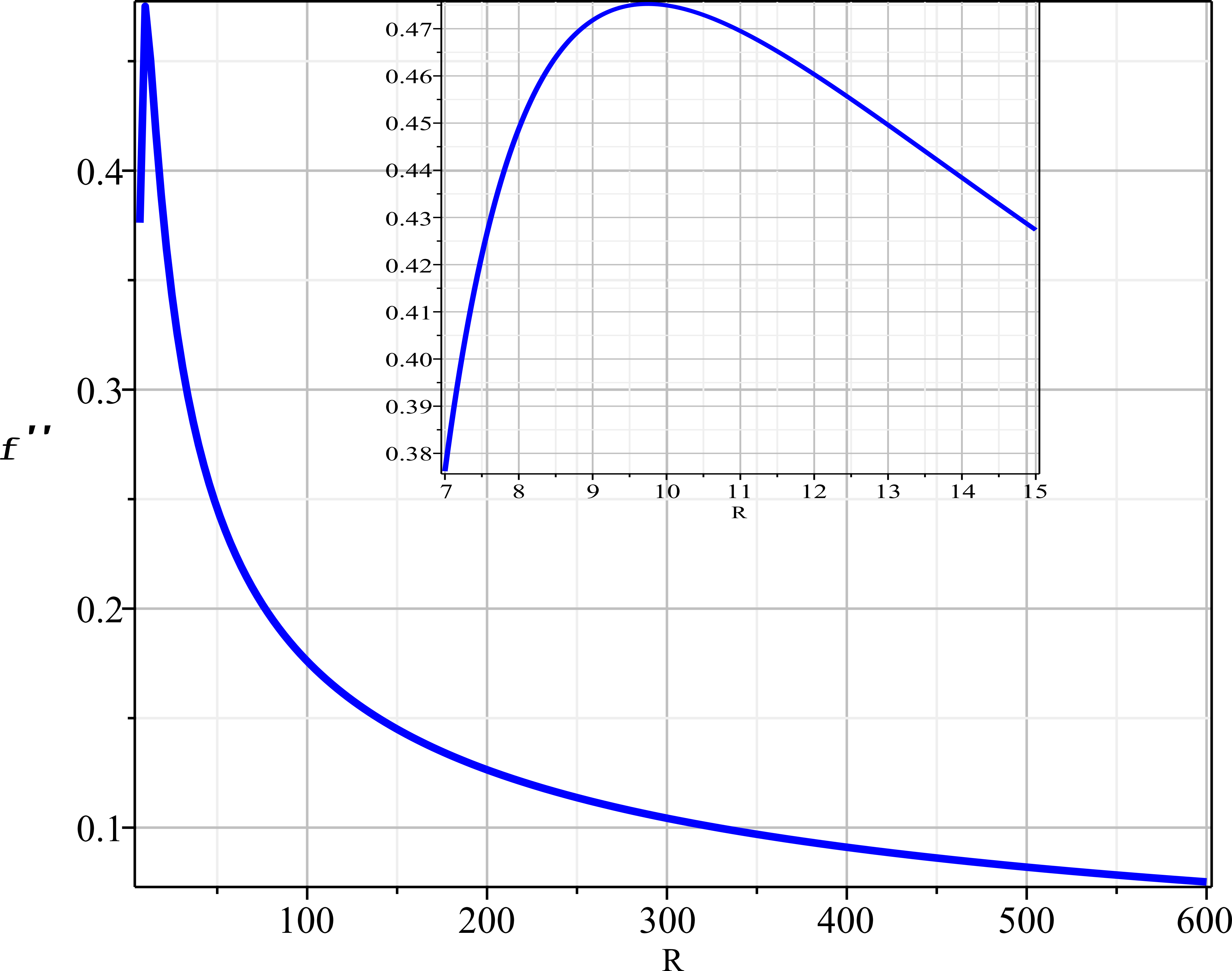}
    \caption{This figure indicates the plot of $f^{\prime \prime}(R)$ (obtained from numerical analysis) versus curvature $R$ at curvature interval $[7,600]$. The input figure demonstrates the behavior of $f^{\prime \prime}(R)$ at low curvature interval $[7,15]$ with a suitable resolution.}
    \label{fzegond(Num)}
\end{figure}

In this part, we proceed with a constant jerk parameter. Limpidly, four options are of interest:
\begin{itemize}
	\item $\rho_{\mathrm{m}}\neq 0$ and $\rho_{\mathrm{em}} \neq 0$;
	\item $\rho_{\mathrm{m}}=0$ and $\rho_{\mathrm{em}} \neq 0$;
	\item $\rho_{\mathrm{m}} \neq 0$ and $\rho_{\mathrm{em}} = 0$;
	\item $\rho_{\mathrm{m}} = 0$ and $\rho_{\mathrm{em}} = 0$.
\end{itemize}
If we keep both densities non-zero, then the analytical solution of equation~(\ref{Ray02}) will be a very complicated case in terms of hypergeometric function. More precisely, besides some hypergeometric functions, there are some complicated analytically unsolvable integrals in terms of hypergeometric functions. The same situation is for the pure electromagnetic case (i.e. $\rho_{\mathrm{m}}=0$ and $\rho_{\mathrm{em}} \neq 0$). It means that the electromagnetic part leads to some analytically unsolvable integrals. Nonetheless, I think that we may do a thing: ``Solving numerically and then fitting the obtained curve with a suitable function in each interval of interest''. This, however, provides an approximation function to $f(R)$ but it helps for observing the manner of the evolution of $f(R)$ and giving a model for $f(R)$-gravity. Between two first options, let us proceed with the generalist case namely $\rho_{\mathrm{m}} \neq 0$ and $\rho_{\mathrm{em}} \neq 0$. It is needless to consider the second option as well since the solving process for both is the same. Solving equation~(\ref{Ray02}) with the conditions
\begin{equation}\begin{split}\label{condi0012}
&f(R=7)=20, \qquad f^{\prime}(R=7)=9.90, \qquad \lambda=\sqrt{3}/2,\\ &\epsilon=-1,\qquad
\rho_{\mathrm{m}}=\rho_{\mathrm{em}}=1/2, \qquad
\left|\epsilon_{1}-\epsilon_{2} \right|=\sqrt{3} \times 10^{-10}
\end{split}\end{equation}
numerically via the Runge-Kutta-Fehlberg 4th order method provides three figures~\ref{f(Num)}--\ref{fzegond(Num)}. The figures are presented in the curvature interval $7$ through $600$. As is observed, both conditions $f^{\prime}(R)>0$ and $f^{\prime \prime}(R)>0$ are satisfied in this large interval. According to figures, it seems that polynomial function or some forms of exponential function are good candidates for fitting the curve. On the other hand, according to the plot of $f^{\prime \prime}(R)>0$, the form of $f(R)$ is at least forth order as
\begin{equation}\label{hamin01}
f(R)=c_{0}+c_{1}R+c_{2}(R-c_{3})^2+c_{4}(R-c_{5})^{3}+c_{6}(R-c_{7})^4,
\end{equation}
where $c_{i}$s are constants.
Now, if we fit the form of $f(R)$ in the curvature interval $10$ to $100$ polynomially with step $1$, we arrive at the form:
\begin{align}\label{curvefitting}
f(R)=&-43.3892391638874 + 7.12150883125319 \; R+0.240246117112700 \; R^2 \nonumber \\& -0.00109345398314376 \; R^3
+3.13314951937404 \times 10^{-6} \; R^4+\mathcal{O}(R^5).
\end{align}
Note that the precision of this work depends upon the curvature interval length. Obviously, fitting in a small interval with small step gives a good approximation for the function.
We started the interval from $10$ instead of $7$ because of ignoring some departures observed in the plot of $f^{\prime \prime}(R)$ (see Fig.~\ref{fzegond(Num)}). Our starting point to numerical analysis was $R=7$ because~(\ref{rhoem002}) and~(\ref{condi0012}) lead to
\begin{equation}\label{rhosadehshodeh}
\rho_{\mathrm{em}}=\frac{(R-4)^{4/3}}{2},
\end{equation}
hence, for having physical behavior, the minimum value of curvature is four.
According to~(\ref{curvefitting}), our approximation type for $f(R)$ looks good for the rest of the interval. It means that at high precision, the form of $f(R)$ would also be a polynomial function.

Two last options namely $\{\rho_{\mathrm{m}} \neq 0 \; \& \; \rho_{\mathrm{em}} = 0\}$ and $\{\rho_{\mathrm{m}} = 0 \; \& \; \rho_{\mathrm{em}} = 0\}$, for FRW case have sufficiently been studied in ref.~\cite{gupta}. The solution for anisotropic one is not so different than FRW's solution. Only the constants will vary a little. Hence we do not discuss this case.\\

\noindent\bahbahmybox{}{\subsubsection{The variable jerk case $j=Q(t)$ ($n=3$)}}\vspace{5mm}

In this part, we study a variable jerk parameter.
Again, like the constant jerk case, here there are four options of interest.\\

\noindent\pqrmybox{red}{\textbf{$\bullet$} \textbf{Pure electromagnetic case ($\rho_{\mathrm{em}} \neq 0$ and $\rho_{\mathrm{m}} = 0$):}}\\

\begin{figure}
	\includegraphics[width=3.5 in]{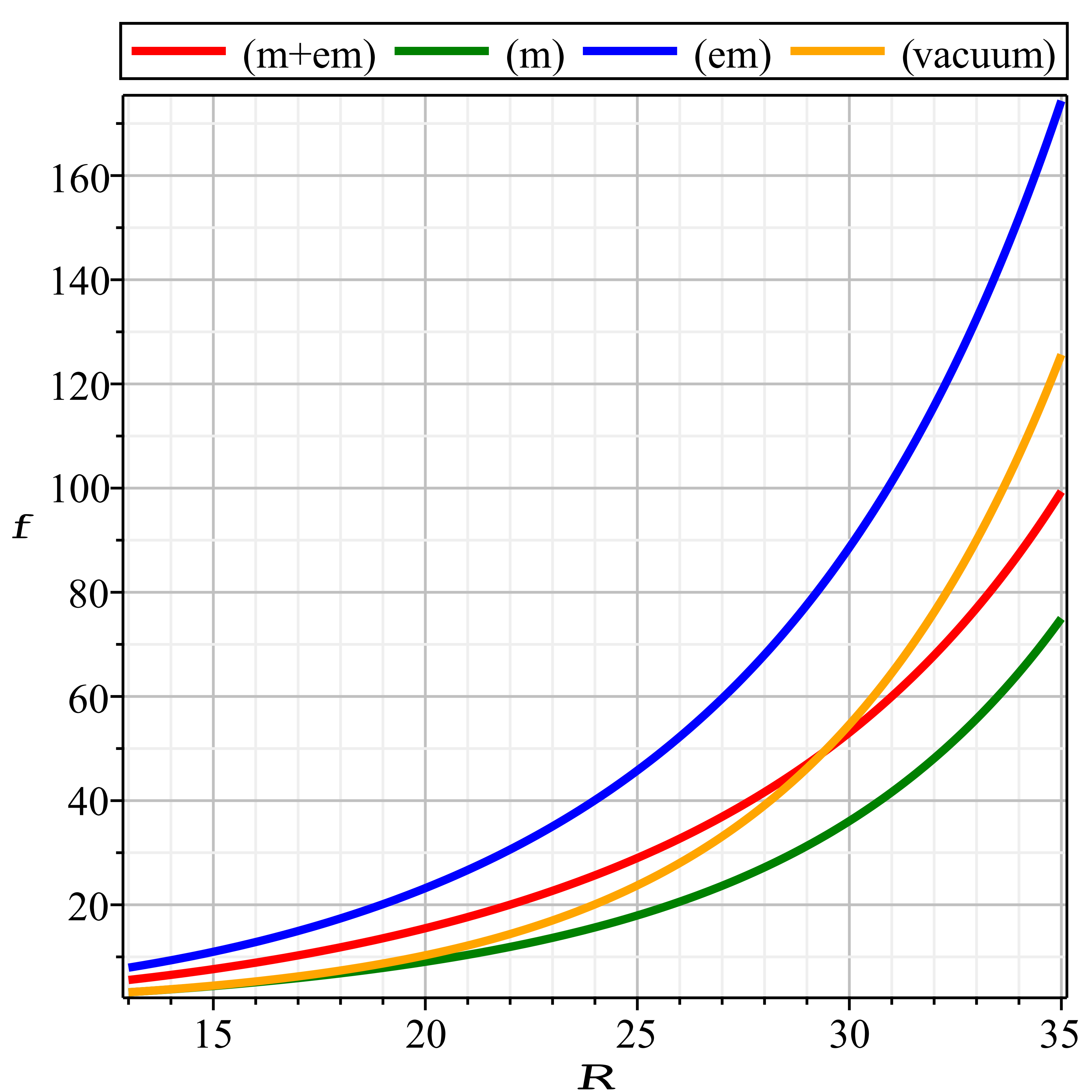}
    \caption{This figure demonstrates the plot of $f(R)$ for four options of interest (1- Pure electromagnetic case (blue color); 2- Pure matter case (green color); 3- Both electromagnetic and matter cases (red color); 4- Vacuum case (orange color)) at curvature interval $[13,35]$.}
    \label{f(4)}
\end{figure}

\begin{figure}
	\includegraphics[width=3.5 in]{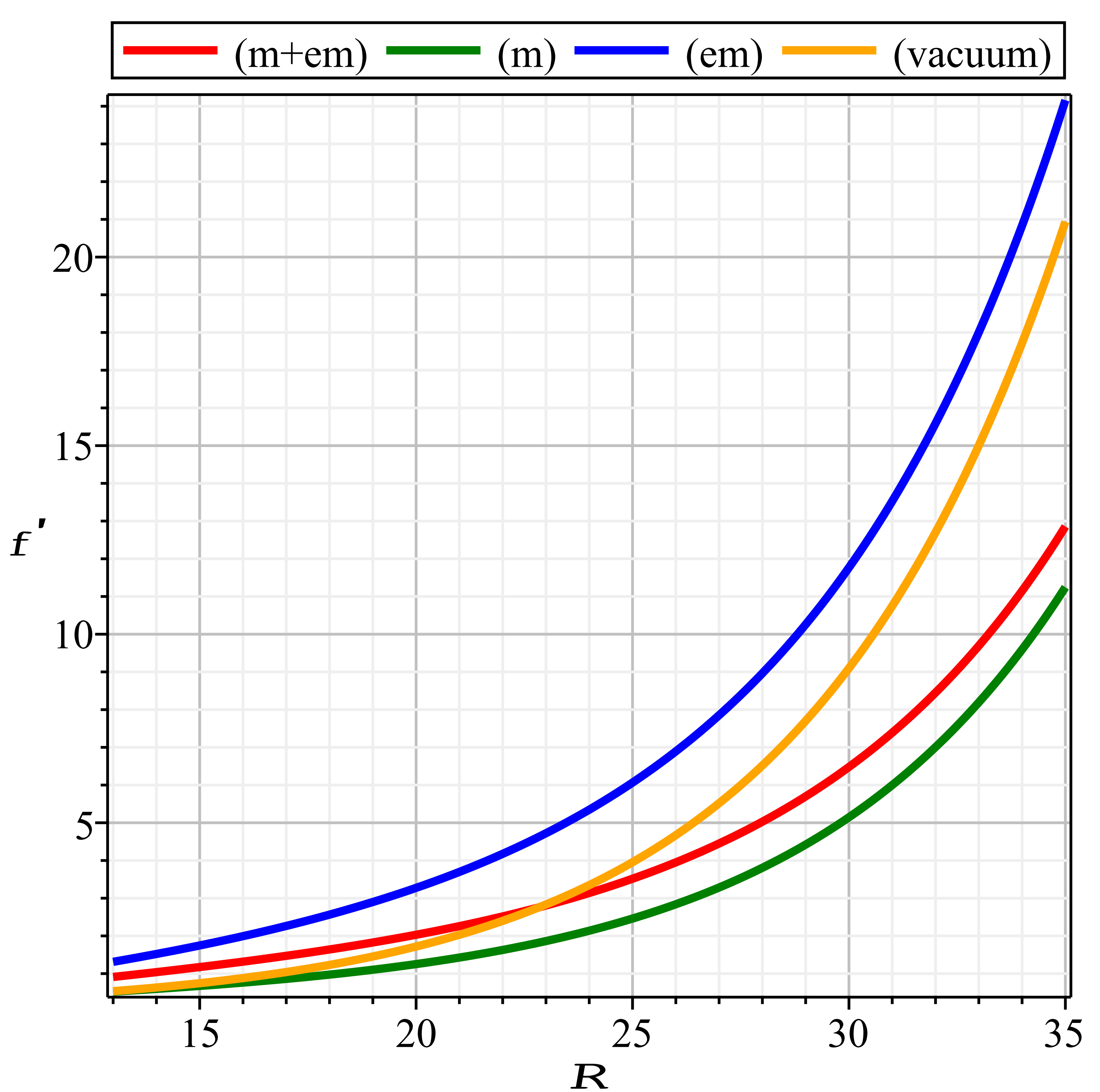}
    \caption{This figure indicates the plot of $f^{\prime}(R)$ for four options of interest (1- Pure electromagnetic case (blue color); 2- Pure matter case (green color); 3- Both electromagnetic and matter cases (red color); 4- Vacuum case (orange color)) at curvature interval $[13,35]$.}
    \label{f(prime-4)}
\end{figure}

\begin{figure}
	\includegraphics[width=3.5 in]{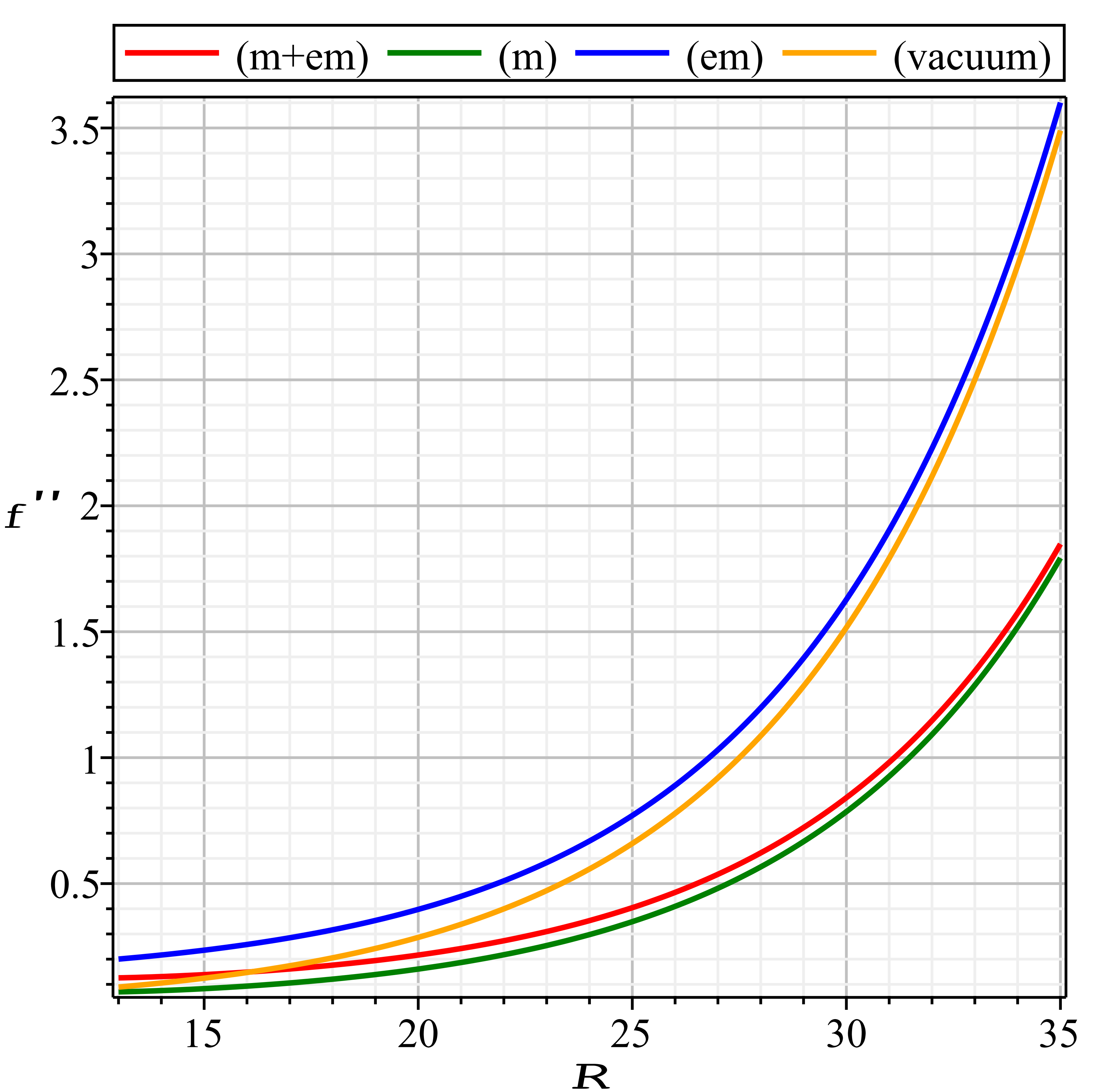}
    \caption{This figure demonstrates the plot of $f^{\prime \prime}(R)$ for four options of interest (1- Pure electromagnetic case (blue color); 2- Pure matter case (green color); 3- Both electromagnetic and matter cases (red color); 4- Vacuum case (orange color)) at curvature interval $[13,35]$.}
    \label{f(prime-prime-4)}
\end{figure}

For this case, the solution to the Raychaudhuri equation is found as
\begin{align}\label{VJ001}
f(R)=2 \rho_{\mathrm{em0}}+ \frac{\rho_{\mathrm{em0}}}{18 \lambda^{4}}\left(R-6\lambda^{2}\right)^{2}+A_{6}\exp \left[\frac{R-6\lambda^{2}}{6\lambda^{2}} \right]
\end{align}
where $A_{6}$ is a constant of integration. This solution holds for both types of extrinsic curvatures in~(\ref{Sol-I})--(\ref{Sol-II}). Note that this solution is obtained for the special case $\epsilon_{1}=\epsilon_{2}=1/3$ namely FRW. For the anisotropic case, the basic equation yields an analytically unsolvable integral. Hence, it should be solved numerically. But since the order of anisotropy of the universe is so little, hence its solution will not so different than this solution. Indeed, the solution~(\ref{VJ001}) can also be regarded as a curve fitted function to anisotropic one as well. This situation is in three cases in what follows as well. According to the solution~(\ref{VJ001}), both validity conditions $f^{\prime}(R)>0$ and $f^{\prime \prime}(R)>0$ are satisfied only by taking $A_{6}>0$. The behaviors of $f$, $f^{\prime}$ and $f^{\prime \prime}$ in the curvature interval $[13,35]$ are presented in Figs.~\ref{f(4)}--\ref{f(prime-prime-4)} with blue color.\\

\noindent\pqrmybox{red}{\textbf{$\bullet$} \textbf{Pure matter case ($\rho_{\mathrm{em}} = 0$ and $\rho_{\mathrm{m}} \neq 0$):}}\\

For this case, the solution to the Raychaudhuri equation for $\epsilon=-1$, which corresponds to~(\ref{Sol-II}), is obtained as
\begin{align}\label{VJ002}
f(R)=&\left(\frac{-3\rho_{\mathrm{m0}}}{\sqrt{6}}\right) \sqrt{\frac{R-12\lambda^{2}}{\lambda^{2}}}
+\left(\frac{\rho_{\mathrm{m0}}}{3\sqrt{6}}\right) \left(\frac{R-12\lambda^{2}}{\lambda^{2}}\right)^{3/2} \nonumber
\\&+\left[A_{7}-\left(\frac{27\rho_{\mathrm{m0}}\sqrt{\pi}}{2 \; e} \right)
\text{erf} \left(\frac{R-12\lambda^{2}}{\sqrt{6}\lambda^{2}}\right) \right]
\exp \left[\frac{R-6\lambda^{2}}{6\lambda^{2}} \right],
\end{align}
where $A_{7}$ is a constant of integration and $e$ is the base of the natural logarithm, $e=2.718281828\cdots$, and ``\,erf\,'' is the error function\footnote{The error function is defined for all complex $u$ by $\text{erf}(u)=\frac{2}{\sqrt{\pi}}\int_{0}^{u}\exp(-t^2)\mathrm{d}t$. The error function is a smooth function which has a simple zero at $u=0$.}.
For the case $\epsilon=+1$, which corresponds to~(\ref{Sol-I}), the corresponded solution yields imaginary form for $f(R)$ which is non-physical, hence we put it aside. As mentioned earlier, the case~(\ref{Sol-I}) has further problems as well.\\
Because the value of error function in every point is in $-1 \leq \text{erf}(x) \leq +1$, hence the last term in~(\ref{VJ002}) is not so strange thing as it can be removed by $A_{7}$. Furthermore, the last term tends to zero as $R$ increases\footnote{We know that:\\ $\lim_{x \to \infty} \left[\left(1-\text{erf}(x) \right)\exp(x) \right]=0$,\\$\lim_{x \to 0}\; \left[\left(1-\text{erf}(x) \right)\exp(x) \right]=1$,\\ because $\text{erf}(0)=0$ and $\text{erf}(\pm \infty)= \pm 1$.} and it means that the effect of this term on the evolution of $f(R)$ is so little such that it may reasonably be ignored.
Indeed, the fluctuations and surplusage produced by the last term are practically petty and insignificant.\\
As is clear from~(\ref{VJ002}), by taking $R>12\lambda^{2}$, both validity conditions ($f^{\prime}>0 \; \& \; f^{\prime \prime}>0$) are satisfied. For example, three figures are presented for this case in which the constants have been taken as $A_{7}=\rho_{\mathrm{m0}}=1$; see green plots in Figs.~\ref{f(4)}--\ref{f(prime-prime-4)}.\\

\noindent\pqrmybox{red}{\textbf{$\bullet$} \textbf{Both electromagnetic and matter cases ($\rho_{\mathrm{em}} \neq 0$ and $\rho_{\mathrm{m}} \neq 0$):}}\\

In general, this case leads to an integral which is analytically unsolvable. But, by choosing $\lambda=+1$ and $\epsilon=-1$, one can arrive at:
\begin{align}\label{VJ003}
f(R)=&2\rho_{\mathrm{em0}} +A_{8} \exp \left[\frac{R-6}{6} \right] +\left(\frac{\rho_{\mathrm{em0}}}{18} \right)\left(R-6 \right)^{2} \nonumber \\&
+\left(\frac{\rho_{\mathrm{m0}}}{108} \right) \left(6R-72 \right)^{3/2}
+\left(\frac{\rho_{\mathrm{m0}}}{2} \right) \sqrt{6R-72}
\nonumber \\&
-\left(\frac{3 \sqrt{\pi}\rho_{\mathrm{m0}}}{2} \right)
\left(\text{erf}\left[\frac{\sqrt{6R-72}}{6} \right] \right)
\exp\left[\frac{R-12}{6} \right],
\end{align}
where $A_{8}$ is a constant of integration. This solution is like the sum of the solutions of two cases which obtained above. Limpidly, the error function comes from the matter part not electromagnetic. Also, the root of existing the powers $1/2$ and $3/2$ backs to matter part while the electromagnetic part produces $+2$ instead.\\
However, the behavior of~(\ref{VJ003}) depends upon the selection of constant parameters, but obviously, in general, both validity conditions for~(\ref{VJ003}) are satisfied by taking $R>12$.
Taking start point $R=13$ and setting $A_{8}=1$ and $\rho_{\mathrm{m0}}=\rho_{\mathrm{em0}}=1/2$, three plots have been presented in the interval $13$ through $35$; see red plots in Figs.~\ref{f(4)}--\ref{f(prime-prime-4)}.\\

\noindent\pqrmybox{red}{\textbf{$\bullet$} \textbf{Vacuum case ($\rho_{\mathrm{em}} = 0$ and $\rho_{\mathrm{m}} = 0$):}}\\

For the vacuum case (i.e. $\rho_{\mathrm{em}} = 0$ \& $\rho_{\mathrm{m}} = 0$) of a varying jerk $j=Q(t)$, a solution to the Raychaudhuri equation would be
\begin{align}\label{vacuumcasej2}
f(R)=A_{9} \exp \left[\frac{R-6\lambda^{2}}{6\lambda^{2}} \right],
\end{align}
where $A_{9}$ is a constant of integration. This solution is arguable from the last three solutions as the term which does not depend upon the densities is this term. Hence, it is the effect of the vacuum case.\\
Clearly, both validity conditions are satisfied only by taking $A_{9}>0$. Unlike three last cases, here, there is no lowers bound for starting the physical values of curvature according to~(\ref{vacuumcasej2}).\\
The related plots to this case have been demonstrated by orange color in Figs.~\ref{f(4)}--\ref{f(prime-prime-4)}.

According to all solutions obtained for $j=Q(t)$ and their Taylor expansions\footnote{We know that $\text{erf}(u)=\frac{2}{\sqrt{\pi}}\sum_{k=0}^{\infty}\frac{(-1)^{k}u^{2k+1}}{k! (2k+1)}$.}, all four $f(R)$s of this case obviously reduce to Einstein's theory at so low curvature. Furthermore, at low curvature, $f$, $f^{\prime}$, and $f^{\prime \prime}$ of four options in view of their amounts, satisfy this relation: $\mathrm{em}$\footnote{Pure electromagnetic case}$>\mathrm{m+em}$\footnote{Both electromagnetic and matter cases }$>\mathrm{vacuum}$\footnote{Vacuum case}$>\mathrm{m}$\footnote{Pure matter case} (see Figs.~\ref{f(4)}--\ref{f(prime-prime-4)}). At high curvature, the arrangement will be changed as $\mathrm{em}>\mathrm{vacuum}>\mathrm{m+em}>\mathrm{m}$.
It means that adding electromagnetic part causes that the amounts of $f$, $f^{\prime}$, and $f^{\prime \prime}$ increase, while adding typical matter/perfect fluid implies that their amounts decrease than vacuum case. In all curvature of interest, $f$, $f^{\prime}$, and $f^{\prime \prime}$ of pure electromagnetic and pure matter cases have the highest and lowest values, respectively. At high curvature, $f$, $f^{\prime}$, and $f^{\prime \prime}$ of vacuum case are greater than the case which contains both matter and electromagnetic while at low curvature, the layout is changed.\\

\noindent\bahbahmybox{}{\subsubsection{A slowly varying jerk parameter}}\vspace{5mm}

In this part, a scenario where the jerk parameter is a slowly varying function of redshift $z$, viz,
\begin{align}\label{svj1}
j(z)=1- \eta_{1} F(z),
\end{align}
where $\eta_{1}$ is a small constant parameter and $F(z)$ is a slowly varying function of the redshift, is considered. Since $j(z)$ varies slowly with respect to $z$, hence it is a good approximate that we take $F(z) \approx F_{0}+F_{1}z$ ($F_{0}$ and $F_{1}$ are constant). Therefore, one may easily get
\begin{align}\label{svj2}
\dot{\Theta} \approx \eta_{2} \Theta^{2},
\end{align}
where $\eta_{2}$ is a constant. Unlike the previous examples, let us here take perfect fluid in general not only pressureless dust fluid (i.e. $p_{\mathrm{m}}=w\rho_{\mathrm{m}}$). For this case, a solution to the Raychaudhuri equation is as follows
\begin{align}\label{svj3}
f(R)=\underbrace{\eta_{3} \; R^{\xi_{1}}+\eta_{4} \; R^{\xi_{2}}}_{\mathrm{vacuum \;\;part}}+\underbrace{\eta_{5}\rho_{\mathrm{em0}} \; R^{4|\beta_{1}|}}_{\mathrm{em-part}}+\underbrace{\eta_{6} \rho_{\mathrm{m0}} \; R^{2(1+w)}}_{\mathrm{m-part}}
\end{align}
where $\eta_{3}$ and $\eta_{4}$ are integration constants and
\begin{align}
\xi_{1} &=\frac{-k_{1}+k_{2}+\sqrt{k^{2}_{1}+k^{2}_{2}-2k_{1}k_{2}+2k_{2}}}{2k_{2}},\\
\xi_{2} &=\frac{-k_{1}+k_{2}-\sqrt{k^{2}_{1}+k^{2}_{2}-2k_{1}k_{2}+2k_{2}}}{2k_{2}}\\
\beta_{1}&=\frac{2}{3}\left(\frac{|\epsilon_{1}-\epsilon_{2}|}{\sqrt{3}}-1 \right),\\
\eta_{5} &=\frac{-2}{\eta_{8} \eta_{9}}\left[8k_{2}w^{2}+\left(4k_{1}+12k_{2} \right)w+4k_{1}+4k_{2}-1\right],\\
\eta_{6} &=\frac{-2}{\eta_{8}\eta_{9}}\left[k_{2}\beta^{2}_{1}+8k_{1}\left|\beta_{1} \right|-8k_{2}\left|\beta_{1} \right|-1 \right],\\
k_{1}&=\frac{1}{\eta_{7}}\left[\eta_{2}+\frac{1}{3}
+\frac{2(\epsilon_{1}-\epsilon_{2})^2}{3} \right],\\
k_{2}&=\frac{-2\eta_{2}}{\eta_{7}},\\
\eta_{7}&=2\eta_{2}+\frac{4}{3}+\frac{2(\epsilon_{1}-\epsilon_{2})^2}{3},\\
\eta_{8}&=8k_{2}w^{2}+4k_{1}w+12wk_{2}+4k_{1}+4k_{2}-1,\\
\eta_{9}&=32k_{2}\beta^{2}_{1}+8k_{1} \left|\beta_{1} \right|-8k_{2} \left|\beta_{1} \right|-1.
\end{align}
Obviously, the powers related to vacuum part namely $\xi_{1}$ and $\xi_{2}$ can take any constant, but it is better we take them positive for the satisfaction of validity conditions. The participation of electromagnetic and perfect fluid are as $R^{4|\beta_{1}|}$ and $R^{2(1+w)}$, respectively. Again like previous cases of study, we observe that unlike the matter part term, the power of curvature arisen from the electromagnetic part is affected by the anisotropic background. Note that $|\beta_{1}| \approx 2/3$ hence the electromagnetic part for this case produces $R^{8/3}$ ($8/3$ is the exact value of FRW background). The special cases namely pressureless dust matter ($w=0$) generates $R^{2}$ which is less than the electromagnetic case. Therefore, pursuant to the previous examples and this one, it seems that the electromagnetic case in most cases of interest, generates the higher powers of curvature than perfect fluid/matter case. It is interesting to note that for the dark energy ($w=-1$) the power $2(1+w)$ would be zero, hence it does not produce any power for the curvature. Both perfect fluid and electromagnetic parts satisfy the validity conditions separately. And as a final point, we mention that Einstein's general relativity theory can be recovered by the terms of the vacuum part or by a matter with the EoS $w=(-1/2)$.\\

\noindent\hrmybox{}{\section{On the enhancement of the method\label{enhance}}}\vspace{5mm}

The presented method has this potential to be applied further. The first generalization is that the inverse road of the method may be adopted to examine a given theory of $f(R)$ according to its emerged outcomes (i.e. behaviors of shear and Hubble, etc.) and comparing them with observational data. But, it seems that in most cases of interest, this application is numerically feasible, not analytically.

According to observational data, the treatment of Hubble, EoS, and etc. are clear. On the other hand, we have Raychaudhuri equation which gives us the form of $f(R)$. Hence, by applying a curve-fitting method to obtained curve from numerical methods, it is feasible to arrive at some forms for $f(R)$ at different stages of the evolution of the universe. This is the second generalization.

Because these tasks are beyond the scope of this paper, hence I do not give an example but pursuant to the first suggestion, I tried for some given forms of $f(R)$ and found that they are only numerically doable.\\

\noindent\hrmybox{}{\section{Conclusions}}\vspace{5mm}

Utilizing Raychaudhuri-based reconstruction strategy suggested by Choudhury et al.~\cite{gupta}, the reconstruction of anisotropic Einstein-Maxwell equation in $1+3$ covariant formalism of $f(R)$-gravity was investigated. The matter part of the problem was assumed to be a non-interacting combination of a perfect fluid and an electromagnetic field. The model has been reconstructed in four interesting modes of evolution. In summary, some of our findings to these modes were as follows:
\begin{enumerate}
	\item A constant deceleration parameter (An accelerating universe):\\
The obtained form for $f(R)$ was as:
$$f(R)= C_{3} R^{l}+\frac{C_{4}}{R^{|k|}}+l_{8}R^{v}+l_{9}R^{n}.$$
The range of powers was given in Fig.~\ref{fig1}. It has been concluded that Einstein's theory does not emerge from this form because of the domains of powers and it is due to the fact that his theory does not give an ever-accelerating universe. The terms $l_{8}R^{v}$ and $l_{9}R^{n}$ in the above form come from matter and electromagnetic parts, respectively. Under the conditions of the problem, one always has $n>v$ and both contribute at the higher orders of curvature which are also reachable via vacuum case.
	\item The constant jerk case $j=1$ mimicking $\Lambda \text{CDM}$ model:\\
Generally, this case is analytically unsolvable. Hence, we proceed using the Runge-Kutta-Fehlberg 4th order method and (functional) curve-fitting method. The outcome for the generalist case up to fifth order polynomial in some curvature interval was as:
\begin{align*}
f(R)=&-43.3892391638874 + 7.12150883125319 \; R \nonumber \\&+0.240246117112700 \; R^2 -0.00109345398314376 \; R^3 \nonumber \\&
+3.13314951937404 \times 10^{-6} \; R^4 +\mathcal{O}(R^5).
\end{align*}
	\item The variable jerk case $j=Q(t)$:\\
For this case, four options were of interest: 1- $\{\rho_{\mathrm{m}}\neq 0 \; \& \; \rho_{\mathrm{em}} \neq 0\}$; 2- $\{\rho_{\mathrm{m}}=0 \; \& \; \rho_{\mathrm{em}} \neq 0\}$; 3- $\{\rho_{\mathrm{m}} \neq 0 \; \& \; \rho_{\mathrm{em}} = 0\}$; 4- $\{\rho_{\mathrm{m}} = 0 \; \& \; \rho_{\mathrm{em}} = 0\}$. The obtained forms to $f(R)$ were respectively as follows:
\begin{align*}
f_{1}(R)=&2\rho_{\mathrm{em0}} +A_{8} \exp \left[\frac{R-6}{6} \right] +\left(\frac{\rho_{\mathrm{em0}}}{18} \right)\left(R-6 \right)^{2} \nonumber \\&
+\left(\frac{\rho_{\mathrm{m0}}}{108} \right) \left(6R-72 \right)^{3/2}
+\left(\frac{\rho_{\mathrm{m0}}}{2} \right) \sqrt{6R-72}
\nonumber \\&
-\left(\frac{3 \sqrt{\pi}\rho_{\mathrm{m0}}}{2} \right)
\left(\text{erf}\left[\frac{\sqrt{6R-72}}{6} \right] \right)
\exp\left[\frac{R-12}{6} \right]; \nonumber \\
f_{2}(R)=&2 \rho_{\mathrm{em0}}+ \frac{\rho_{\mathrm{em0}}}{18 \lambda^{4}}\left(R-6\lambda^{2}\right)^{2}+A_{6}\exp \left[\frac{R-6\lambda^{2}}{6\lambda^{2}} \right]; \nonumber \\
f_{3}(R)=&\left(\frac{-3\rho_{\mathrm{m0}}}{\sqrt{6}}\right) \sqrt{\frac{R-12\lambda^{2}}{\lambda^{2}}}
+\left(\frac{\rho_{\mathrm{m0}}}{3\sqrt{6}}\right) \left(\frac{R-12\lambda^{2}}{\lambda^{2}}\right)^{3/2} \nonumber
\\&+\left[A_{7}-\left(\frac{27\rho_{\mathrm{m0}}\sqrt{\pi}}{2e} \right)
\text{erf} \left(\frac{R-12\lambda^{2}}{\sqrt{6}\lambda^{2}}\right) \right]
\exp \left[\frac{R-6\lambda^{2}}{6\lambda^{2}} \right]; \nonumber \\
f_{4}(R)=&A_{9} \exp \left[\frac{R-6\lambda^{2}}{6\lambda^{2}} \right].
\end{align*}
Pursuant to these forms, reconstruction via variable jerk led to an exponential function of $f(R)$, $\exp(R-R_{0})$, and the contribution of matter and electromagnetic parts appeared as $\{(R-R_{0})^{3/2} \; \& \; (R-R_{0})^{1/2} \; \& \; \text{erf}(R-R_{0}) \times \exp(R-R_{0})\}$ and $\{(R-R_{0})^2\}$, respectively.
All these interesting obtained forms at low curvature tend to Einstein's theory of gravity.
	\item A slowly varying jerk parameter with redshift:\\
For this case, the form of $f(R)$ found out as
\begin{align*}
f(R)=\eta_{3} \; R^{\xi_{1}}+\eta_{4} \; R^{\xi_{2}}+\eta_{5}\rho_{\mathrm{em0}} \; R^{4|\beta_{1}|}+\eta_{6} \rho_{\mathrm{m0}} \; R^{2(1+w)}.
\end{align*}
The first two terms come from vacuum part while the third and fourth terms arise from the electromagnetic and perfect fluid, respectively. Hence, the participation of the electromagnetic part for FRW is as $R^{8/3}$. For the pressureless dust matter, the power of curvature will be $+2$ and for the dark energy, the power of curvature is zero.

The validity conditions for all $f(R)$s obtained from four modes of evolution were satisfied (in some cases entirely and in others at special intervals or under specific conditions).
\end{enumerate}

There is a interesting common property among all cases studied in this paper:\\
\textit{Unlike the perfect fluid/matter part, the power of curvature produced by the electromagnetic part is affected by anisotropic property of background. Furthermore, the power of curvature supplied by electromagnetic part is higher than matter/perfect fluid part.} For example, for FRW case, some of our findings were as follows:
\begin{itemize}
	\item For constant deceleration parameter: $\mathcal{P}_{\mathrm{em}}>2$ and $\mathcal{P}_{\mathrm{m}}>3/2$;
	\item For variable jerk parameter with time: $\mathcal{P}_{\mathrm{em}}=2$ and $\mathcal{P}_{\mathrm{m}}=3/2 \; \& \; 1/2$;
	\item For slowly varying jerk parameter with redshift: $\mathcal{P}_{\mathrm{em}}=8/3$ and $\mathcal{P}_{\mathrm{m}}=2$,
\end{itemize}
where $\mathcal{P}_{\mathrm{em}}$ and $\mathcal{P}_{\mathrm{m}}$ refer to the powers of curvatures of the electromagnetic and matter parts, respectively.

Finally, some discussions about the enhancement of the method were done.

\section*{\noindent\goldmybox{red}{\vspace{3mm} Acknowledgments \vspace{3mm}}}

This work has been supported financially by Research Institute for Astronomy $\&$ Astrophysics of Maragha (RIAAM) under research project No. 1/6275-20. \\

\hrule \hrule \hrule \hrule \hrule \hrule

\end{document}